\title{decadal2020_Garcia}
\documentclass[11pt]{article} % Paper type (a4paper, usletter or legal) and font size (10, 11 or 12)

\usepackage[letterpaper,margin=1.00in]{geometry}
\usepackage{mdframed}
\usepackage{deluxetable}
\usepackage{comment}

\usepackage{sidecap}  % caption to the side of the fig
\usepackage{graphics,graphicx}
\graphicspath{{figs/}{other-documents/CVs/}}     % Path to where the figures live
\usepackage{wrapfig,framed} % Allows wrapping text around figures
\usepackage{amssymb}
\usepackage{natbib,natbibspacing,url,multicol}
\bibpunct{(}{)}{;}{a}{}{,}
\usepackage[font={footnotesize}]{caption}
\usepackage[dvipsnames]{xcolor}
\usepackage{tabularx}
\usepackage{sidecap}
\usepackage{footnote}
\usepackage{lscape}
\usepackage{xcolor,colortbl}
\usepackage{amsmath}
\usepackage{colortbl,hhline}

\usepackage{enumitem}
\newlist{compactitem}{itemize}{3} % 3 is max-depth
\setlist[compactitem]{label=\textbullet, nosep}
\def\gx{GX~339--4}
\def\cygx1{Cyg~X--1}

\def\msun{$M_{\odot}$}

\definecolor{silvia_blue}{RGB}{117, 178, 219}
\definecolor{silvia_green}{RGB}{191, 219, 145}
\definecolor{silvia_darkblue}{RGB}{48, 79, 158}
\definecolor{silvia_gray}{RGB}{152, 152, 153}

\usepackage{charter} % Charter font for main content

\usepackage[cm]{sfmath} % Better font for the equations

\usepackage{amssymb,amsmath} % Math packages
\usepackage{multicol} % Required for the three-column layout of the document
\usepackage{url} % Clickable links
\usepackage{enumitem} % Reduces the amount of space within and between lists with [noitemsep,nolistsep]
\usepackage{marvosym} % Required for the use of symbols
\usepackage[T1]{fontenc} % Use 8-bit encoding that has 256 glyphs
\usepackage{datetime} % Required for defining a custom date style
\newdateformat{mydate}{\monthname[\THEMONTH] \THEYEAR} % Set a custom date format

\usepackage[colorlinks = true,
            linkcolor = red,
            urlcolor  = blue,
            citecolor = blue,
            anchorcolor = blue]{hyperref}

\usepackage[nottoc,notlot,notlof]{tocbibind}  % This makes the references appear in the TOC

\usepackage[final]{pdfpages}

\usepackage{fancyhdr} % Required to define custom headers/footers
\newcommand{\HorRule}[1]{\noindent\rule{\linewidth}{#1}} % Creates a horizontal rule

\newcommand{\NewsletterName}[1]{ % Newsletter title
\begin{center}
\vspace{30pt}
\LARGE \usefont{T1}{fvs}{}{n} % Use the Bera Sans Bold font
#1
\vspace{10pt}
\end{center}	
\par \normalsize \normalfont}

\newcommand{\JournalIssue}[1]{ % Date and issue number at the top of the newsletter
\hfill \textsc{Astro 2020 Decadal Science White Paper, \mydate \today} % Right-aligned date and issue number
\par \normalsize \normalfont}

\usepackage{multirow}
\usepackage{bigstrut}
\newcolumntype{P}[1]{>{\raggedright}p{#1}}
\newcolumntype{M}[1]{>{\raggedright}m{#1}}
\newlength{\rowA}
\usepackage{titlesec}
\titleformat{\section}
  {\bfseries\large\color{Blue}}
  {\thesection}{0.4em}{}
\titleformat{\subsection}
  {\sffamily\large\bfseries\color{Blue}}
  {\thesubsection}{0.4em}{}

\begin{document}

\JournalIssue{1} % Issue number
\NewsletterName{Probing the Black Hole Engine with Measurements of the Relativistic X-ray Reflection Component}
\vspace{-5pt}
%\NewsAuthor{
%}
\noindent\HorRule{3pt} \\[-0.75\baselineskip] % Thick horizontal rule
\HorRule{1pt} % Thin horizontal rule

%----------------------------------------------------------------------------------------
%       MAIN NEWS ITEM
%----------------------------------------------------------------------------------------

\thispagestyle{empty}
\hypersetup{linkcolor=black}
%\tableofcontents
\vspace{20pt}
{\large

\noindent Thematic Areas:

\begin{itemize}
  \item Formation and evolution of compact objects
  \item Galaxy Evolution
\end{itemize}

\vspace{10pt}
\noindent Principal Author:
\vspace{10pt}

{\bf \textcolor{Blue}{Javier A. Garc\'ia}}$^{1,2}$ 

{\it $^1$California Institute of Technology}

{\it javier@caltech.edu}

{\it +1-626-395-6609}

\vspace{20pt}
\noindent Co-authors (alphabetical):
\vspace{10pt}

{\textcolor{Blue}{
Matteo Bachetti$^{1,3}$,
David R. Ballantyne$^4$,
Laura Brenneman$^5$,
Murray Brightman$^1$,
Riley M. Connors$^1$,
Thomas Dauser$^2$,
Andrew Fabian$^6$,
Felix Fuerst$^7$,
Poshak Gandhi$^8$,
Nikita Kamraj$^1$,
Erin Kara$^{9,10,11}$,
Kristin Madsen$^1$,
Jon M. Miller$^{12}$,
Michael Nowak$^{13}$,
Michael L. Parker$^7$,
Christopher Reynolds$^6$,
James Steiner$^{11}$,
Daniel Stern$^{14}$,
Corbin Taylor$^9$,
John Tomsick$^{15}$,
Dominic Walton$^6$,
J\"orn Wilms$^2$, \&
Abderahmen Zoghbi$^{12}$
}

\vspace{10pt}
{\it $^2$Dr. Karl-Remeis Observatory, Germany},
{\it $^3$INAF-Osservatorio Astronomico di Cagliari, Italy},
{\it $^4$Georgia Tech University},
{\it $^5$Smithsonian Astrophysical Observatory},
{\it $^6$University of Cambridge, UK},
{\it $^7$European Space Agency, ESAC, Spain},
{\it $^8$University of Southampton, UK},
{\it $^9$University of Maryland College Park},
{\it $^{10}$NASA Goddard Space Flight Center},
{\it $^{11}$Massachusetts Institute of Technology},
{\it $^{12}$University of Michigan},
{\it $^{13}$Washington University in St. Louis},
{\it $^{14}$NASA Jet Propulsion Lab},
{\it $^{15}$University of California, Berkeley}
}

\vfill
\begin{flushright}
%\sffamily
{\it ``Nothing can be loved or hated unless it is first understood"}

---Leonardo da Vinci
\end{flushright}

\hypersetup{linkcolor=red}
\newpage
\setcounter{page}{1}

%----------------------------------------------------------------------------------------
\normalsize
\vspace{-0.2cm}
\section{Introduction}\label{sec:intro}
\vspace{-0.3cm}

Over thirty years ago, Soviet physicist and Nobel laureate Vitaly Ginzburg
wrote: {\it ``If the cosmological problem is the number one problem of
astronomy, then problem number two should be the problem of black holes"}
\citep{gin85}.  This powerful statement, also cited by \cite{mcc09} in their
Astro 2010 Decadal White Paper, is equally relevant today.  In physics, black
holes represent one of the best laboratories to test general relativity.  In
astrophysics, stellar-mass black holes are the endpoints of stellar evolution,
providing the unique opportunity to study the gravitational collapse of matter
into a single point laying beyond the light-trapping event horizon. Conversely,
the origin and growth of their supermassive counterparts in active galactic
nuclei (AGN), believed to be important drivers of galactic evolution, remains
an open question.  Today, the field continues to make momentous strides, with
the detection of gravitational waves (GW) from LIGO/Virgo
\citep[e.g.,][]{abb16,abb17b}, and with the Event Horizon Telescope (EHT)
expected to deliver the first images of the the black hole in the center of our
Galaxy \citep[e.g.,][]{lu18}.

Since the first discovery of a black hole, Cygnus~X-1, X-ray astronomy has
offered a unique probe into the nature of compact objects.  Energetic radiation
reflected from the accretion disk around a black hole produces emission lines
that can be distorted by Doppler and gravitational shifts \citep{fab89}. These
effects were first confirmed with the observation of the relativistically
broadened Fe K emission in the AGN spectrum of MCG--6-30-15 \citep{tan95}. In
the last two decades, X-ray spectroscopy has proven to be a powerful tool for
the estimation of black hole spin and several other physical parameters in
dozens of AGN and black hole X-ray binaries \citep[BHXBs;][]{rey13}.

\vspace{-0.1cm}

\begin{mdframed}
\vspace{-0.1cm}
{\it
In this White Paper, we discuss the observational and theoretical challenges
expected in the exploration, discovery, and study of astrophysical black holes
in the next decade. We focus on the case of accreting black holes and their
electromagnetic signatures, with particular emphasis on the measurement
of the relativistic reflection component in their X-ray spectra.
}
\vspace{-0.1cm}
\end{mdframed}

%----------------------------------------------------------------------------------------
\vspace{-0.2cm}
\section{X-ray Reflection Spectroscopy}\label{sec:ref} 
\vspace{-0.3cm}

Prior to the detection of GW from merging black holes, accreting systems
(either AGN or BHXB) have provided the best opportunity to study the properties
of black holes. Furthermore, given that the basics of accretion physics are
independent of the mass of the central object, and in contrast to GW, the same
techniques can in principle be applied to systems with any black hole mass,
e.g., a BHXB with masses in the 1.4--100\,\msun~range, or AGN with masses in
the $10^5-10^9$\,\msun~range \footnote{In some cases, such as X-ray reflection
spectroscopy, the techniques are also applicable to a binary system where the
compact object is a neutron star \citep[e.g.,][]{cac08,lud18}.} (i.e., LIGO \&
EHT will not probe the same wide range of masses).  In many of these systems,
infalling gas forms a flat rotating structure known as an accretion disk, with
the matter spiraling slowly toward the center on Keplerian orbits. Near the
black hole, thermal emission from the optically-thick disk peaks in the X-ray
band ($kT \sim 1$\,keV) for BHXB, and in the ultraviolet ($kT \sim 10$\,eV) for
AGN \citep{sha73}. Additionally, a non-thermal power-law component of emission
is ubiquitous, arising through Comptonization by much hotter electrons ($kT
\sim 100$\,keV) in an optically-thin region that is referred to as the {\it
corona}.
%

%...................................................................................
\begin{figure}
\begin{framed}
\includegraphics[width=\linewidth]{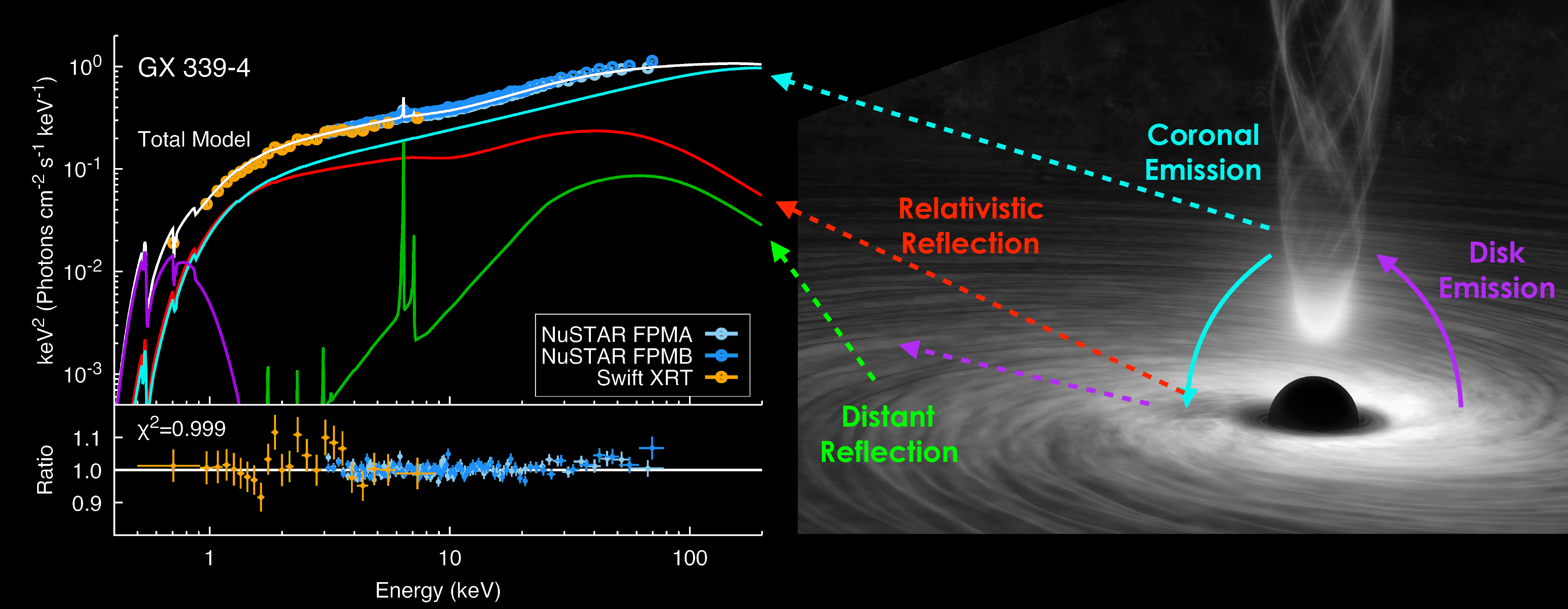}
\vspace{-0.2in}
\caption{({\it left:}) The {\it Swift} and {\it NuSTAR} spectrum of the BHXB GX~339$-$4
during its 2017 outburst, fitted with a Compton continuum (blue); relativistic
reflection (red); distant reflection (green); and thermal disk emission
(violet). The lower panel shows the fit residuals (Garc\'{\i}a et al. in prep.)
({\it right:}) Schematic representation for the origin of the spectral
components in an accreting black hole.  The disk's thermal emission (violet) is
Compton scattered into a power-law (blue) by electrons in a hot and compact
corona. A fraction of this component illuminates the disk thereby generating
the relativistic reflection component (red arrows), as well as a distant
reflection component (green). These models are commonly used to constrain the
black hole spin, among other important parameters. Adapted from original artwork by
NASA/JPL-Caltech/R. Hurt (IPAC).
} \label{fig:cartoon}
\end{framed}
\vspace{-20pt}
\end{figure}
%...................................................................................

The hard coronal radiation illuminates the relatively cold accretion disk and
creates a spectral component typically referred to as the {\it reflection
spectrum}, which is composed of a forest of fluorescent lines, edges and
related features \citep{ros93,gar10}. These reprocessed X-rays leave the disk
carrying information about the physical composition and condition of the matter
in the strong field near the black hole. The most prominent feature is the
fluorescent Fe K complex of emission lines at 6--7\,keV. The line profiles are
grossly distorted in the strong gravity regime by Doppler effects, light
bending and gravitational redshift \citep{fab89,dau13}.
Figure~\ref{fig:cartoon} shows a schematic view of these components. By
modeling the shape of the Fe K profile and the entire reflection spectrum, much
can be deduced about matter near the black hole and about the black hole
itself, including its {\it spin}, or angular momentum. An alternative method to
measure the spin of stellar-mass black holes is based on the fitting of the
thermal-disk emission \citep{zha97}.  Profoundly, spin and mass are the two
physical quantities that completely define an astronomical black hole, with the
spin providing an indirect record of its formation mechanism and growth history
\citep{moderski98,vol05}.  Additionally, reflection spectrum fitting can be
complemented by timing analysis of X-ray reverberation signatures (lags that
result from path length differences between the direct and reprocessed X-rays),
which can be used to further probe the properties of the inner disk, hot
corona, and central compact object \citep{utt14}.

Relativistically-broadened Fe K lines have by now been observed in the spectra
of most well-studied BHXBs and a large fraction of AGN. \cite{bre13b} and
\cite{rey14} list $\sim$20 AGN with estimates of black hole spin, while the Fe
line has been detected in many additional AGN. For BHXBs, \cite{rey14} lists
spin estimates for 14 of the 18 black holes cataloged by \cite{mcc06}. In the
last couple of years, 11 new black hole candidate systems have been discovered,
from which at least 5 exhibit broad Fe line emission: Swift~J1858.6$-$0814
\citep{lud18b}, MAXI~J1820+070 \citep{kar19}, MAXI~J1535$-$571 \citep{xu18a},
Swift~J1658.2$-$4242 \citep{xu18b}, and MAXI~J1631$-$479 \citep{miy18}.
%

%----------------------------------------------------------------------------------------
\vspace{-0.2cm}
\section{Open Questions in Black Hole Astrophysics for the Next Decade}\label{sec:goals}
\vspace{-0.3cm}

\noindent\underline{\bf The Spin of Black Holes Across Mass and Redshift.}
Obtaining accurate estimates of spin is a key goal in black-hole astrophysics.
In the case of supermassive black holes, the distribution of spin versus mass
can be used to discriminate among different formation scenarios and
evolutionary paths, such as hierarchical \citep{vol05} or chaotic \citep{kin08}
growth via galaxy mergers, or steady growth via standard radiatively-efficient
accretion \citep{bar72}.  While current observations suggest that broad Fe K
lines are very common in the local universe, it is unclear if they were also
common at earlier times, such as at the peak of AGN activity at $z\sim$ 0.5--4.
Current observational data has been pushed to the limit to achieve high-$z$
spin estimates using techniques leveraging gravitational lensing
\citep{rei14,reym14,cha18}, Bayesian fitting \citep{bar18}, and stacking
\citep{wal15}.  The current census of spin measurements for local AGN is
limited in the size of the sample ($\sim 30$ objects), precision (some
measurements are only upper or lower limits), and accuracy, as observational
biases and systematic uncertainties are not yet well understood \citep[e.g.,
see ][for a recent review]{rey19}.

For stellar-mass black holes, accurate spin measurements are crucial for
understanding their formation and evolutionary histories
\citep[e.g.,][]{fra15}, as well as exploring the suggested correlation between
jet power and black hole spin \citep{nar12,ste13,mcc14,che16}. Accurate spins
and masses for samples of stellar-mass black holes can be used to discern
between different axion models, such as via the search for evidence of black
hole superradiance \citep{bri15}, a process through which angular momentum and
energy are predicted to be extracted from the compact object by large numbers
of bosons that populate gravitationally bound states \citep{bar17}.

\noindent\underline{\bf Understanding the Physics of Accretion onto Black
Holes.} As a BHXB cycles between quiescence and its Eddington luminosity, it
exhibits a wide range of behaviors that includes AU-scale steady jets,
parsec-scale ballistic jets, X-ray quasi-periodic oscillations (QPOs) with
frequencies spanning $0.01-450$\,Hz, and distinct ``hard" and ``soft"
spectral/timing states \citep{fen04, rem06}; all of which may be tied to the
spin of the black hole.

A critical assumption in measuring black hole spin is that the inner edge of
the accretion disk is located at the innermost stable circular orbit (ISCO).
This assumption has been firmly validated for accretion disk dominated soft
states \citep[e.g.,][]{ste10,zhu12}, but for Comptonization dominated hard
states, it is a matter of dispute. Measurements of the inner-disk radius
$R_\mathrm{in}$ with reflection models and timing techniques appear to be in
disagreement by orders of magnitude for the same source \citep[e.g.,
\gx;][]{gar15,bas15,dem16}.  The question of disk truncation has important
ramifications for measurements of black hole spin, given the strong degeneracy
between inner radius and spin that arises due to their relations to the
strength of gravitational redshift \citep[e.g.,][]{fab14}, which limits the
confidence with which either parameter can be obtained.  Improvement in data
quality is thus crucial to resolve this controversy, as higher signal and
superior spectral resolution will overcome current model degeneracies.
Moreover, physical parameters obtained from reflection spectroscopy analysis
can be used to inform our understanding of other aspects of accretion physics,
such as the importance of the vertical structure as opposed to a razor-thin
disk \citep{tay18,tay2018b}, the possibility of disk warps or misalignment
\citep[e.g.,][]{tom14,mid16,mil18}, or the presence of disk winds and outflows
\cite[e.g.,][]{mil15,par18}.  Thus, continuous monitoring of accreting black
holes is needed in order to provide a comprehensive picture of the dynamical
evolution of accretion disks throughout different state transitions.

\noindent\underline{\bf The Detailed Microphysics of Accretion Disks.} The
prevalence of super-solar iron abundances is an unexpected result from
reflection spectroscopy studies of BHXBs and AGN \citep{gar18a}, which calls
into question the accuracy of spin estimates given the strong correlation
between spin and Fe abundance \citep[e.g.,][]{rey12,ste12}.  This systematic
tendency for very large Fe abundances has motivated the revision of X-ray
reflection models. In particular, recent calculations for densities above the
traditional values ($n_e \sim 10^{15}$~cm$^{-3}$), show a strong excess of the
reflected continuum at soft energies ($\lesssim 1$\,keV), due to the
enhancement of free-free heating in the atmosphere of the disk, which increases
with increasing density \citep{bal04,gar16b}. Moreover, at sufficiently high
densities ($\gtrsim 10^{18}$~cm$^{-3}$), plasma effects such as atomic
screening and modifications to the nuclear potential become important
\citep[e.g.,][]{dep18}.  Such densities are not only plausible but indeed
expected in accretion disks around black holes, particularly for those with
stellar masses \citep{sch13}.

Application of these new high-density reflection models---which are still under
development---to study both AGN \citep[e.g., IRAS~13224$-$3809, Mrk~1044, and
Mrk~509;][]{jia18,mal18,gar19} and BHXBs \citep[e.g., Cyg~X-1 and
GX~339--4;][]{tom18,jia19} suggests that high-density effects are significant,
and must be correctly considered in order to properly interpret the observed
X-ray spectrum. In all these cases, the iron abundance recovered by the model
was significantly decreased relative to results obtained with lower-density
disk reflection models.  The analysis of a statistically significant sample of
sources with improved models and state-of-the-art atomic data promises to
reveal exciting new details of the microphysics of accretion disks around black
holes.

\noindent\underline{\bf The Origin of the Corona.} The geometry, location, and
even the origin of the source that illuminates the disk are still largely
unknown. The photons are thought to originate from a Comptonizing hot corona
\citep[e.g.,][]{haa93b,dov97,zdz03}, which may be associated with the base of a
jet \citep[e.g.,][]{mat92,mar05}. Recent observations show evidence for a
thermal cutoff in the spectra of several AGN \citep[e.g.,][]{fab14,loh15},
suggesting that the corona is close to the pair-production limit.  The electron
temperature and optical depth of the corona can both be accurately constrained
by modeling the reflection spectrum, which indicates that the corona responds
dramatically to changes in luminosity \citep{gar15,kar19}. Assumptions made on
the properties of the corona have a direct impact on the shape of the
reflection spectrum.  Meanwhile, there exists abundant observational evidence
for X-ray reverberation originating from the disk's innermost region, just
outside the event horizon \citep[e.g.,][]{zog10,kar16}, thus confirming that
the broad features observed in the spectrum are due to relativistic reflection.
The detailed modeling of these observables will provide new ways to estimate
the properties of the corona.
%

%----------------------------------------------------------------------------------------
\vspace{-0.2cm}
\section{Further Advances and Efforts Required}\label{sec:other}
\vspace{-0.3cm}

In order to significantly expand our understanding of the physics of black
holes and the behavior of accretion disks in strong gravity, further
development in reflection models is necessary, along with the improvement of
observational capabilities in the next decade. This includes the calculation
and measurement of atomic parameters for high density plasmas, allowing the
extension of reflection models to densities above $10^{18}$\,cm$^{-3}$.  Future
work should also explore alternative coronal illumination patterns
\citep[e.g.,][]{wil16}, as well as incorporate detailed calculations of disk
thermal emission \citep[e.g.,][]{bal01,ros07}, ionization gradients
\citep[e.g.,][]{svo12,kam19}, and realistic density profiles.  It will be
necessary to understand and accurately assess systematic errors, such as disk
thickness \citep{tay18,tay2018b}, Comptonization of reflection features by the
corona \citep{wil15,ste17}, and realistic temperature profiles in the disk
(hardening factors) \citep{davis18}.  Implementing the results from GR-MHD
simulations \citep[e.g.,][]{kinch16,kinch18} into reflection models would
provide a new means of testing accretion theory predictions and hypotheses
about plasmas in strong gravity. Finally, new modeling methodologies, such as
the simultaneous modeling of lag-energy and lag-frequency spectra
\citep[e.g.,][]{cha16,cab17}, or direct analysis of the cross-spectrum
\citep[e.g.,][]{mas18,bachetti18}, would allow us to use data more effectively,
better guaranteeing self-consistency and overcoming model degeneracies.
%

%----------------------------------------------------------------------------------------
\vspace{-0.2cm}
\section{Concluding Remarks}\label{sec:conc}
\vspace{-0.3cm}

Achieving a deeper understanding of accreting black holes requires the next
leap in our observational capabilities (summarized in Table~\ref{tab:sum}).
Future missions flying micro-calorimeters such as {\it XRISM} \citep{tas18},
{\it Athena} \citep{nan13}, and {\it Lynx} \citep{oze18}, will provide
unprecedented spectral resolution in the X-ray band. The characterization of
the reflection spectrum also requires high sensitivity in the 10--100\,keV band.
Concept probe missions such as {\it HEX-P} \citep{mad18}, {\it STROBE-X}
\citep{ray18}, and {\it eXTP} \citep{zha16}, will provide such capabilities;
while observations of AGN at high redshift or in crowded fields can only
be achieved with a superior angular resolution such as that in the concept
mission {\it AXIS} \citep{mus18}.
%

%----------------------------------------------------------------------------------------
\newcolumntype{s}{>{\columncolor{silvia_green}} p{2.5cm}}
\newcolumntype{t}{>{\columncolor{silvia_blue}} p{3.5cm}}
\newcolumntype{g}{>{\columncolor{silvia_gray}} p{2.4cm}}
\setlength{\arrayrulewidth}{0.3mm}
\arrayrulecolor{black}
%\begin{landscape}

\begin{table}[!hb]
%\vspace{-0.5cm}
\center
\caption{Summary of the Science Goals for Black Hole Astrophysics with X-ray Reflection Modeling}
\label{tab:sum}
\begin{scriptsize}
\begin{tabular}{|s|t|p{4.0cm}|p{4.8cm}|} % could be |c|M{2cm}|
\hline
\rowcolor{silvia_darkblue}
\textcolor{white}{Science Goals} & \textcolor{white}{Key Questions} & \textcolor{white}{Methodology}&\textcolor{white}{Desired Observational Capabilities} \\
%
%----------------------
\hline
\vspace{0.3ex}
{
\textbf{Goal 1:}

Measure the distribution of supermassive black hole spins in the local Universe}
&

\vspace{0.3ex}
\begin{itemize}[nosep,leftmargin=1em,labelwidth=*,align=left]
\item Were the supermassive black holes in the local Universe grown via galaxy
merges or steady growth?
\end{itemize} &

\vspace{0.3ex}
{Robust spin estimates ($\lesssim 10$\% uncertainty) for a statistically significant
sample of AGN (min~50; ideally 80--100), using X-ray reflection spectroscopy} &

\vspace{0.3ex}
\begin{itemize}[nosep,leftmargin=1em,labelwidth=*,align=left]
\item Energy band: $\sim 1-150$\,keV
\item Effective area: $\gtrsim 10^4$\,cm$^2$ @ 6\,keV
\item Energy resolution: $< 150$\,eV @ 6\,keV
\item Low background
\vspace{5pt}
\end{itemize} \\

%----------------------
\hline
\vspace{0.3ex}
\textbf{Goal 2:}

Estimate the fraction of highly spinning black holes at the peak of AGN
activity ($z\sim 0.4-5$)
&

\vspace{0.3ex}
\begin{itemize}[nosep,leftmargin=1em,labelwidth=*,align=left]
\item How do supermassive black holes acquire their mass and angular momentum?
\end{itemize} &

\vspace{0.3ex}
Detection of broad Fe K lines and estimate spins for high-$z$ AGN 
($\lesssim$20\% uncertainty) &

\vspace{0.3ex}
\begin{itemize}[nosep,leftmargin=1em,labelwidth=*,align=left]
\item Energy band: $\sim 0.1-15$\,keV)
\item Effective area: $\gtrsim 10^4$\,cm$^2$ @ 1\,keV
\item Energy resolution: $<50$\,eV @ 1\,keV
\item Low background
\item High angular resolution: $<5"$
\vspace{5pt}
\end{itemize}
\\

%----------------------
\hline
\vspace{0.3ex}
\textbf{Goal 3:}

Measure the distribution of BHXB spins &

\vspace{0.3ex}
\begin{itemize}[nosep,leftmargin=1em,labelwidth=*,align=left]
\item How do stellar-mass black holes acquire their spin and mass?
\item Are X-rays probing a different population of black holes from those measured with LIGO/Virgo?
\item Are jets powered by black hole spins?
\vspace{-5pt}
\end{itemize}

%\vspace{0.5ex}
%* Can axions be produced via superradiance? &

&

\vspace{0.3ex}
Robust spin estimates ($\lesssim 10$\% uncertainty) for a statistically 
significant sample of BHXB
(min 30; ideally 50--100), using X-ray reflection spectroscopy. Validate results with other
techniques when possible &

\vspace{0.3ex}
\begin{itemize}[nosep,leftmargin=1em,labelwidth=*,align=left]
\item Energy band: $\sim 1-150$\,keV
\item Effective area: $\gtrsim 10^3$\,cm$^2$ @ 6\,keV
\item Energy resolution: $<150$\,eV @ 6\,keV
\item High count rate capabilities (no pile-up) to a few Crab
\item Fast ToO response: $<24-48$\,hrs
\end{itemize}
\\

%----------------------
\hline
\vspace{0.3ex}
\textbf{Goal 4:}

Understand the dynamics of black hole accretion physics &

\vspace{0.3ex}
\begin{itemize}[nosep,leftmargin=1em,labelwidth=*,align=left]
\item Is the accretion disk significantly truncated in BHXBs during the hard state?
\item Which are the main physical quantities that drive the state changes? 
\end{itemize}
&

\vspace{0.3ex}
Measure the level of disk truncation to high precision ($5-10$\% uncertainty) 
and its evolution during the outburst of several BHXBs with X-ray reflection &

\vspace{0.3ex}
\begin{itemize}[nosep,leftmargin=1em,labelwidth=*,align=left]
\item Energy band: $\sim 1-150$\,keV
\item Effective area: $\gtrsim 10^3$\,cm$^2$ @ 6\,keV
\item Energy resolution: $<100$\,eV @ 6\,keV
\item High count rate capabilities (no pile-up) to a few Crab
\item Fast ToO response: $<24-48$\,hrs
\item Monitoring capabilities: daily observations during weeks/months 
\vspace{-5pt}
\end{itemize}
\\

%----------------------
\hline
\vspace{0.3ex}
\textbf{Goal 5:}

Study the detailed microphysics of accretion disks &

\vspace{0.3ex}
\begin{itemize}[nosep,leftmargin=1em,labelwidth=*,align=left]
\item Is iron significantly over-abundant in most accretion disks?
\item What's the origin of the soft-excess in AGN? 
\end{itemize}
&

\vspace{0.3ex}
Determine the density of the accretion disk in a strong gravitational field for
a sample of BHXBs and AGN using state-of-the-art reflection models &

\vspace{0.3ex}
\begin{itemize}[nosep,leftmargin=1em,labelwidth=*,align=left]
\item Energy band: $\sim 0.1-150$\,keV
\item Effective area: $\gtrsim 10^4$\,cm$^2$ @ 6\,keV
\item Energy resolution: $<150$\,eV @ 6\,keV
\item High count rate capabilities (only for BHXBs) to a few Crab
\item Polarization capabilities
\vspace{-5pt}
\end{itemize}
\\

%----------------------
\hline
\vspace{1pt}
\textbf{Goal 6:}

Investigate the origin, geometry, and behavior of the X-ray
corona in accreting black holes &

\vspace{0.3ex}
\begin{itemize}[nosep,leftmargin=1em,labelwidth=*,align=left]
\item What are the properties of the corona?
\item How does the corona evolve throughout state transitions?
\end{itemize}
&

\vspace{0.5ex}
Measurements of the electron temperature and optical depth of the coronal
emission (via modeling the X-ray continuum), as well as the illumination
profile in the accretion disk (via modeling the X-ray reflection spectrum)
\vspace{5pt}
&

\vspace{0.5ex}
\begin{itemize}[nosep,leftmargin=1em,labelwidth=*,align=left]
\item Energy band: $\gtrsim 200$\,keV
\item Effective area: $\gtrsim 10^4$\,cm$^2$ @ 6\,keV
\item Effective area: $\gtrsim 10^2$\,cm$^2$ @ 100\,keV
\item Low background
\item High count rate capabilities (only for BHXBs) to a few Crab
\end{itemize}
\\

%----------------------
\cline{1-4}
\end{tabular}
\end{scriptsize}
\end{table}

%----------------------------------------------------------------------------------------
\newpage

\bibliography{my-references}

%----------------------------------------------------------------------------------------
\end{document}